\documentclass{iau} 
\usepackage{graphicx}
\usepackage{natbib}
\usepackage{comment}
\usepackage{subfigure}

\title[3D AMR simulations of G2 as an outflow] 
{3D AMR simulations of G2 as an outflow}

\author[A. Ballone et al.]   
{A. Ballone$^{1,2}$, M. Schartmann$^{1,2,3}$, A. Burkert$^{1,2,4}$, S. Gillessen$^2$, P.M. Plewa$^2$, O. Pfuhl$^2$, R. Genzel$^2$, F. Eisenhauer$^2$, T. Ott$^2$, E.M. George$^2$ and M. Habibi$^2$}

\affiliation{$^1$University Observatory Munich, Scheinerstra{\ss}e 1, D-81679 M{\"u}nchen, Germany\\
$^2$ Max-Planck-Institute for extraterrestrial Physics, Giessenbachstra{\ss}e 1, \\ D-85741 Garching bei M{\"u}nchen, Germany\\
$^3$Centre for Astrophysics and Supercomputing, Swinburne University of Technology,\\ P.O. Box 218, Hawthorn, Victoria 3122, Australia\\
$^4$ Max-Planck-Fellow}

\pubyear{2016}
\volume{322}  
\jname{The Multi-Messenger Astrophysics of the Galactic Centre}
\editors{S. Longmore, G. Bicknell \& R. Crocker, eds.}
\begin{document}

\maketitle

\begin{abstract}
We study the evolution of G2 in a \textit{Compact Source Scenario}, where G2 is the outflow from a low-mass central star moving on the observed orbit. This is done through 3D AMR simulations of the hydrodynamic interaction of G2 with the surrounding hot accretion flow. A comparison with observations is done by means of mock position-velocity (PV) diagrams. We found that a massive ($\dot{M}_\mathrm{w}=5\times 10^{-7} \;M_{\odot} \; \mathrm{yr^{-1}}$) and slow ($v_\mathrm{w}=50 \;\mathrm{km\; s^{-1}}$) outflow can reproduce G2's properties. A faster outflow ($v_\mathrm{w}=400 \;\mathrm{km\; s^{-1}}$) might also be able to explain the material that seems to follow G2 on the same orbit.
\end{abstract}

\firstsection 
\section{Observations}

In year 2012, \citet{Gillessen12} discovered a small cloud, later named ``G2'', at few thousands Schwarzschild radii from SgrA*. G2 has both a dust component, visible in the near infrared $L'$ band, and a gaseous component, visible in Br$\gamma$ and other recombination lines. The cloud lies on a very eccentric orbit ($e \approx 0.98$) and reached its pericenter in early 2014, with a distance from SgrA* of $\approx 2400$ Schwarzschild radii ($R_S$). The line emission shows an increasing spatial extent and a broadening in the velocity space, interpreted as tidal stretching of the cloud by the tidal field of SgrA* \citep{Gillessen13a,Gillessen13b}. Observed position-velocity (PV) diagrams and Br$\gamma$ maps also show the presence of a tail (G2t), following G2 on roughly the same orbit \citep{Pfuhl15}.

\begin{figure}[h]
\begin{center}
 \includegraphics[scale=0.25]{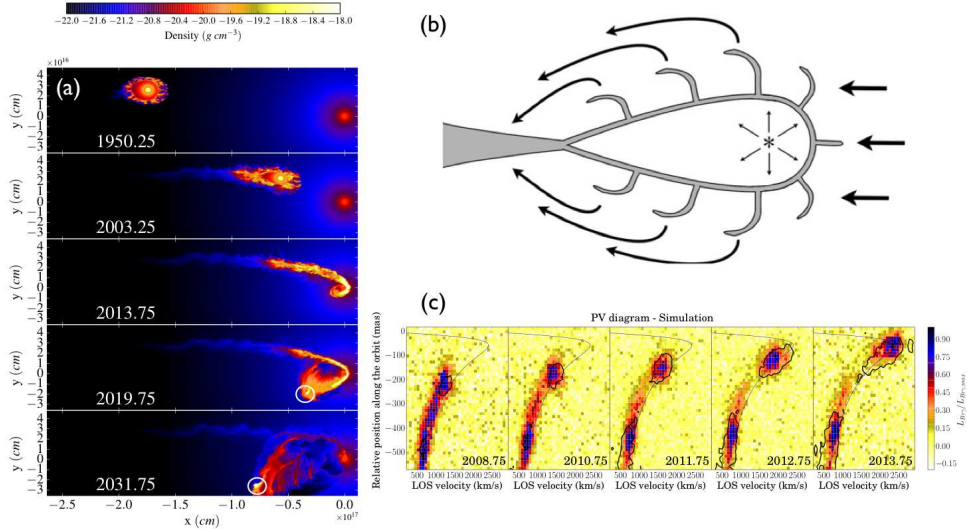} 
 \caption{a) Density distribution for the low velocity model described in the text. The white circles show the outflow reforming after pericenter. b) sketch of the outflow's evolution at late time, for the high velocity model described in the text. c) Simulated PV diagrams for the high velocity model. The black contours mark the observed G2 and G2t.}
   \label{fig1}
\end{center}
\end{figure}
\section{Simulations of massive outflows}

We performed 3D adaptive mesh refinement (AMR) hydrodynamic simulations with the code PLUTO to understand whether G2's gas component can be the outflow of a central source. We start the source at apocenter and let it move on the best-fit orbit derived by \citet{Gillessen13b}. We include an idealized ADAF-like atmosphere and the gravitational field of SgrA* ($M_{BH} = 4.31\times 10^6 \;M_{\odot}$). The outflow initially expands with a spherical shape and stalls due to pressure balance. Close to pericenter, the ram pressure of the atmosphere and the tidal force of the SMBH increase. This leads to the stripping of the Rayleigh-Taylor fingers of shocked outflow material and to the stretching of the cloud (see Fig. 1a). 
We compare to observations by constructing mock position-velocity (PV) diagrams.
For relatively high mass-loss rates ($\dot{M}_\mathrm{w}=5\times 10^{-7} \;M_{\odot} \; \mathrm{yr^{-1}}$) and low velocity ($50 \;\mathrm{km\; s^{-1}}$), the shocked outflow material can reasonably reproduce G2 in the observed PV diagrams, as in the case of the \textit{Diffuse Cloud Scenario} (see conference contribution by Marc Schartmann). For same mass-loss rate, but higher velocity ($400 \;\mathrm{km\; s^{-1}}$), the external ram pressure is more effective and the stripped material accumulates in the trailing region, forming a long tail (see Fig. 1b). The simulated PV diagrams, in the latter case, result in a bimodal emission \citep[see Fig. 1c and][for a detailed discussion]{Ballone16}. In the latter scenario, G2 would be produced by the leading stagnation shock of the outflow and G2t by the long tail of stripped material.

\section{Does G2 contain a compact source?}

As in the case of previous studies \citep{Ballone13, DeColle14}, our best matching parameters seem to suggest a massive outflow, typical of young stellar objects such as T Tauri stars \citep[as already suggested by][see also the conference contribution by Michal Zaja{\v c}ek]{Scoville13}. A T Tauri star is also appealing, since it could naturally explain the presence of dust embedded in G2. The nature of G2 remains an open question; however, in the case of a compact source, we should be able to observe a decoupling between the source and the previous outflow (i.e., between the dust and gas components) in the next 5-10 years (see Fig. 1a). A new G2 should reform around the source, later on.

\end{document}